\documentstyle[12pt,epsfig]{article}

\begin{document}

\title{
\font\fortssbx=cmssbx10 scaled \magstep2
\hbox to \hsize{
\hskip.35in \raise.1in
\hbox{\fortssbx University of Wisconsin - Madison}
\hfill$\vcenter{\normalsize\hbox{\bf MADPH-00-1174}
                       \hbox{May 2000}
                       \hbox{\hfil}}$}
High Energy Neutrinos from Gamma Ray Bursts:\\[-1ex]
Event Rates in Neutrino Telescopes}

\author{J. Alvarez-Mu\~niz, F. Halzen, D.W. Hooper\\
\it Department of Physics, University of Wisconsin, Madison, WI 53706}

\maketitle

\begin{abstract}
Following Waxman and Bahcall \cite{waxman} we calculate
the event rate, energy and
zenith angle dependence of neutrinos produced in the fireball model
of gamma ray bursts (GRB). We emphasize the primary importance of
i) burst-to-burst fluctuations and ii) absorption of the neutrinos
in the Earth. From the astronomical point of view, we draw attention
to the sensitivity of neutrino measurements to the boost Lorentz
factor of the fireball $\Gamma$, which
is central to the fireball model, and only indirectly determined by
follow-up observations. Fluctuations result in single bursts emitting
multiple neutrinos, making it possible to determine the flavor
composition of a beam observed after a baseline of thousands of
Megaparsecs.
\end{abstract}

{\bf PACS number(s):} 96.40.Tv, 98.70.Rz, 98.70.Sa


\section{Introduction}

The origin of GRB is one of the most fascinating outstanding
problems in astronomy. Their observed energy injection in the
Universe is sufficiently large to possibly resolve another long-standing
puzzle:
the origin of the highest energy cosmic
rays\,\cite{waxmanprime,steckerGRB}.
Mounting evidence suggests that GRB emission is produced by a
relativistically expanding fireball, energized by a process involving
neutron stars or black holes\,\cite{piran}.
In the early stages the fireball, its
radiation trapped by the very large optical depth, cannot emit photons
efficiently. The fireball's kinetic energy is therefore dissipated until
it becomes optically thin --- a scenario that can explain the observed
energy and time scales of GRBs, provided the bulk Lorentz factor of the
expanding flow, $\Gamma$, is $\ge 10^2-10^3$.

Protons accelerated in shocks in the expanding fireball interact with
photons to produce charged pions, the parents of high-energy neutrinos
\cite{waxman,vietri}.
Assuming that particles accelerated in the GRB sources produce the
observed
cosmic rays above the ``ankle" of the energy spectrum near
$3\times 10^{18}$~eV, one derives that the average single burst
produces only $\sim 10^{-2}$ neutrino events in a
high energy neutrino telescope with $1\rm~km^2$ effective area.
Although the expected rate is therefore low, the neutrino signal
should be relatively easy to observe provided the detector is
large enough: GRB neutrinos will have a hard spectrum extending
well beyond the background from atmospheric neutrinos and,
even more important, the high-energy GRB neutrino events should
coincide with observed GRB photon events within a narrow time window.

In this paper we calculate the experimental signatures of GRB
in a kilometer-scale neutrino detector such as the proposed
IceCube \cite{icecube}.
We emphasize the importance of taking into account burst-to-burst
fluctuations\, \cite{hooper} as well as absorption of the neutrino signal
in the Earth for both event rates and experimental signatures.
Both effects produce additional and striking signatures with
discriminating sensitivity to the value of the bulk Lorentz factor whose
value is only indirectly inferred from other astronomical
observations \cite{piran,dermer}.

The observation of GRB neutrinos over a cosmological baseline has
scientific potential beyond testing the ``best-buy" fireball model:
the observations can test with unmatched precision special relativity
and the equivalence principle, and study oscillating neutrino flavors
over the ultimate baseline of $z \simeq 1$ \cite{waxman}.

\section{Calculation of GRB Neutrino Rates and Signatures}

In calculating the event rates and experimental signatures of
GRB neutrinos in a high energy neutrino telescope we follow
the model of Waxman and Bahcall \cite{waxman} as implemented by Halzen
and Hooper \cite{hooper}. 
We have normalized the neutrino flux to the energy rate injected
in the Universe needed to explain the observed cosmic ray (CR) spectrum
above $10^{19}$ eV,
$\dot E_{\rm CR}=4\times 10^{44}~{\rm ergs~Mpc^{-3}~yr^{-1}}$.
This energy rate was calculated in reference \cite{waxmanCR}, assuming a
cosmological distribution of sources and taking into account CR propagation
in the Cosmic Microwave Background Radiation (CMBR). The value quoted above
corresponds to the ``low redshift'' ($z<1$) energy generation rate of CRs.
Note that, because of the absorption of highest energy cosmic rays by CMBR
photons, one could further increase $\dot E_{\rm CR}$ without directly
affecting their observed flux. Waxman and Bahcall \cite{wblimit}
have calculated an upper limit to the cosmic rate production rate in the
whole Universe, assuming the energy generation rate evolves rapidly with
redshift following the luminosity density evolution of quasi-stellar
sources. They obtain an upper limit which is $\sim 3~{\dot E_{\rm CR}}$.
It is interesting however to mention that $\dot E_{\rm CR}$
is comparable to that produced in $\gamma$-rays by cosmological GRBs
(which are not expected to be absorbed by the intervening backgrounds
since the typical photon energy is below 1 MeV). Assuming the
efficiency with which electrons (which ultimately produce the observed
photons by synchrotron radiation) and protons is the same inside the
GRB fireball, the value of $\dot E_{\rm CR}$ quoted above might well be
closer to the actual value. For this, and other reasons, it is anyway
unlikely that our calculations are accurate to better than a factor 3 or so.
Moreover our neutrino event rate calculation might be conservative in
this respect.

The neutrino flux is given by

\begin{eqnarray}
{dN_\nu\over dE_\nu}=\cases {{A \over E_B} {1\over E_\nu}
~;~E_\nu<E_B \cr
{A\over E_\nu^2}~;~E_\nu>E_B}
\label{nuflux}
\end{eqnarray}
where $A$ is a normalization constant that is determined from
energy considerations  as explained above and in reference \cite{hooper}.
The observed neutrino rates are calculated by folding this generic
flux over the distributions of individual bursts in distance, $f(z)$,
energy, $f'(E_{\rm GRB})$ and boost factor, $f''(\Gamma)$:
\begin{equation}
N_\nu \propto \int\int\int\int
{dN_\nu\over dE_\nu}(E_{\rm GRB}, z, \Gamma, E_\nu) \,
P(E_\nu)~f(z) \,
f'(E_{\rm GRB}) \, f''(\Gamma)~dE_{\rm GRB}~dz~d\Gamma~dE_\nu\,.
\label{eq:folding}
\end{equation}
where $E_{\rm GRB}$ is the energy emitted by a particular GRB,
$z$ its redshift and $\Gamma$ the boost factor.
$P(E_\nu)$ represents the efficiency of detecting a neutrino
of energy $E_\nu$.
The first two distributions can be modelled after observations:
a cosmological distribution in distance, and
an energy distribution which assumes that ten percent of GRB
produce more energy than average by a factor of ten, and one
percent by a factor of 100\,\cite{piran}.

Most important however are  the fluctuations in the $\Gamma$ factor
around its average phenomenological value of $10^2$--$10^3$.
The fluctuations in $\Gamma$ affect the efficiency for producing
pions in the $p -\gamma$ collisions in the fireball as $\Gamma^{-4}$
\cite{waxman}, as well as the break energy, $E_B$, which varies as
$\Gamma^{2}$. Unfortunately the distribution in $\Gamma$ cannot even
be guessed at. Nevertheless, it is critical in making quantitative
predictions\,\cite{hooper}. The physics is clear. In GRBs, high
luminosities are emitted over short times, therefore the large photon density
would render GRB opaque unless $\Gamma$ is very large. Only
transparent sources with large boost factors emit photons.
They are however relatively weak neutrino sources because the
actual photon target density in the fireball is diluted by the
large Lorentz factor. An even moderately reduced value of $\Gamma$
will produce a prolific neutrino source.

We remind the reader that the results obtained from
Eq.~\ref{eq:folding} are at variance with the neutrino
rate obtained by multiplying the average rate per burst by the
number of bursts. Even neglecting the all important fluctuations
in $\Gamma$, there is no such concept as an average GRB. E.g.\
for $\Gamma=300$, the correct computation of Eq.~\ref{eq:folding}
yields a rate of $\sim 75$ events per km$^2$ and year, roughly an order
of magnitude larger than the prediction obtained by neglecting the
observed burst-to-burst fluctuations in distance and energy. Another
consequence of fluctuations is that the signal is dominated by a few
very bright bursts, which greatly simplifies their detection.

As pointed out above, the average neutrino energy varies with the
square of boost factor and therefore the calculated event rates,
especially their dependence on $\Gamma$, is strongly affected by
the fact that higher energy neutrinos are preferentially absorbed
in the Earth before reaching the detector\,\cite{gaisser}. Also this
effect has been neglected in all previous calculations. For instance, the
75 events just mentioned are reduced by a factor 3 by absorption. One
should on the other hand remember that by oscillations, a large fraction of
$\nu_{\mu}$ can oscillate into
$\nu_{\tau}$ which penetrate the Earth \cite{saltzberg}.
It is in this context important to
realize that a kilometer-scale detector
such as IceCube can measure the energy of the neutrinos. Therefore,
signal events can be separated from the low energy atmospheric background
by energy measurement, which allows the identifications of neutrinos from
all directions and not just in the hemisphere where they pass through
the Earth.

\section{Results}

\begin{figure}[h]
\begin{center}
\mbox{\epsfig{file=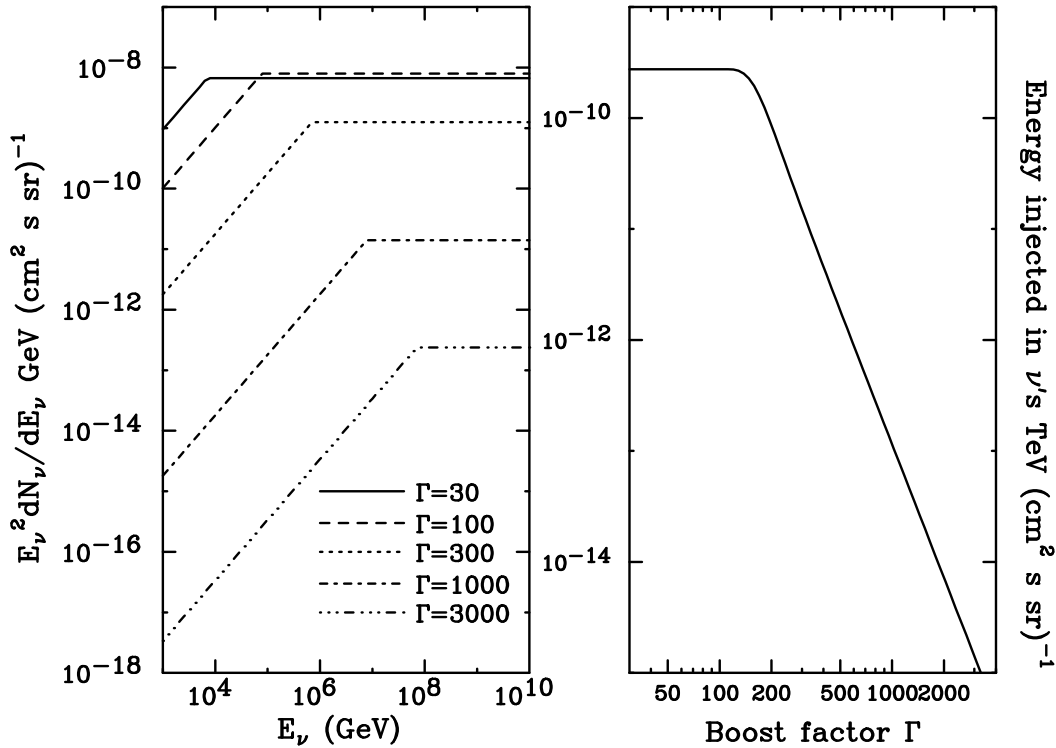,height=10cm}}
\end{center}
\caption{Left: $\nu_\mu+\bar\nu_\mu$ fluxes from GRB
for different values of the Lorentz factor
$\Gamma$ assuming GRBs are responsible
for the observed cosmic ray spectrum above $10^{19}$ eV.
Right: Energy injected in GRB neutrinos as a function of
the boost factor $\Gamma$ of the fireball.}
\label{grbfluxes}
\end{figure}

We calculate the flux of neutrinos from GRB in the fireball model
following reference\ \cite{hooper}. The number of protons in the fireball
is fixed by the assumption that they are the source of the ultra high energy
cosmic rays
above $10^{19}$ eV. The results are shown in Figure \ref{grbfluxes} which
shows the $\nu_\mu+\bar\nu_\mu$ flux
from GRBs for different values of $\Gamma$. The fluxes have been
multiplied by $E_\nu^2$ so that they represent
the energy emitted in the form of neutrinos. Notice the variation of the
break in the spectrum $E_B$ as $\Gamma^2$. Also, for
values of $\Gamma$ below $\sim$ 100 the fireball becomes opaque to
protons and at this point the total amount of energy available for neutrino
production is converted. The neutrino flux, which roughly scales as
$\Gamma^{-4}$, saturates and no longer grows with decreasing boost factor;
see Figure \ref{grbfluxes}.

As usual \cite{gaisser}, we calculate the event rate per GRB by folding the
neutrino flux with the probability of detecting a muon produced in a
$\nu_\mu$ or $\bar\nu_\mu$ interaction. Fluctuations in distance $z$ and
total GRB energy, or fluency, $E_{\rm GRB}$ are accounted for by a Monte
Carlo simulation we have developed for this purpose.
We simulate a large number of GRBs at different zenith angles
assuming an isotropic distribution and subsequently obtain the event rate
per year by assuming 1000 GRBs/year. Absorption of the neutrinos in the
Earth prior to reaching the detector is taken into account using the
density profile of reference \cite{gandhi}. It is also important to
implement the fact that above the detector there is a limited
column density of atmosphere and ice available for neutrino detection.

In Table \ref{numuevents} we separately show, for different $\Gamma$'s,
the event rate of upgoing, i.e. neutrinos that cross the Earth before
interacting near the detector, and downgoing neutrinos.
The first two columns show for comparison the number of events when
neither absorption nor the limited amount of target above the detector are
taken into account. It is clear that
both effects play an important role in obtaining the correct event
rate. This is not surprising:
the Earth becomes opaque to neutrinos of energy around 100 TeV,
and the muon range
exceeds the $\sim$ 2 km vertical depth of the IceCube detector at energies
around $1$ TeV.

\begin{table}
\begin{center}
\begin{tabular}{|c|c|c|c|c|} \hline
& \multicolumn{2}{c|}{~No absorption~} &
\multicolumn{2}{c|}{~Absorption~}\\
\cline{2-5}
\raisebox{2.5ex}[0pt]{~Events/(${\rm km^2}$~yr) in $2\pi$
sr~}&~Downgoing~~&~Upgoing~~&~Downgoing~
~&~Upgoing~~\\ \hline\hline
~$\Gamma=100$~&~1133~&~1112~&~476~&~600~ \\ \hline
~$\Gamma=300$~&~38~&~38~&~13~&~14~ \\ \hline
~$\Gamma=1000 $~&~0.14~&~0.15~&~$4.2\times 10^{-2}$~&~$2.8\times 10^{-2}$~ \\ \hline
\end{tabular}
\end{center}
\caption[] {$\nu_\mu+\bar\nu_\mu$ events $(\rm km^{-2}~yr^{-1})$.
Only fluctuations in distance and energy are taken into account.
\label{numuevents}}
\end{table}

\begin{figure}[h]
\begin{center}
\mbox{\epsfig{file=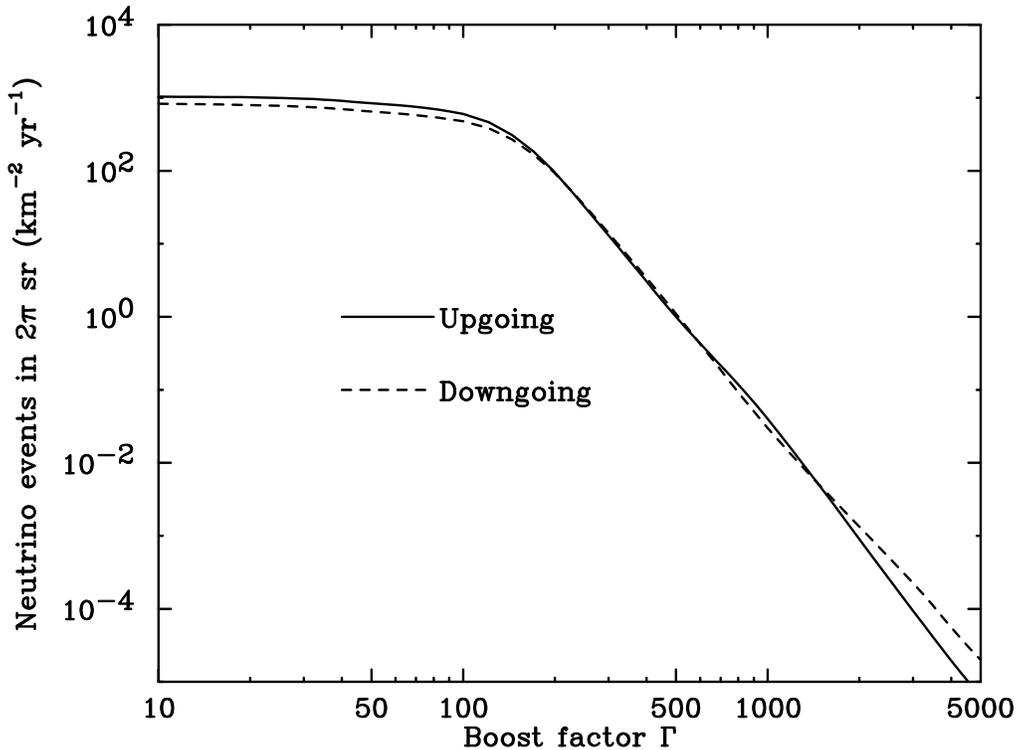,height=10cm}}
\end{center}
\caption{$\nu_\mu+\bar\nu_\mu$ event rate per ${\rm km^2~yr}$
as a function of the boost factor
$\Gamma$ taking into account fluctuations in distance and the
GRB fluency. Absorption in the Earth
and the limited target above the detector are taken
into account.}
\label{grbgamma}
\end{figure}

The dependence of the number of events on the $\Gamma$ factor
is shown in Figure \ref{grbgamma}. Two competing effects determine the
shape of the curve. The
event rate decreases with increasing $\Gamma$ following the dependence of
the neutrino flux which varies as $\Gamma^{-4}$. This decrease is
partially offset because higher energy neutrinos resulting from larger
boost factors are more efficiently detected. For low values of $\Gamma$,
below about 100, the saturation of the total energy available for neutrino
production is seen. On the other side of the $\Gamma$ range,
for large values of $\Gamma$,
the spectrum is very flat ($\sim E^{-1}$) up to $\sim 70$ PeV
where the absorption by the Earth dominates. This reduces the event
rate of upgoing events for values of $\Gamma$ above 1000 as can be seen
in Figure \ref{grbgamma}.
In the end the downgoing and upgoing event
rates are similar except for large values of $\Gamma$.
It is important to keep in mind that although downgoing neutrinos are not
affected by absorption, their detection is limited by the column
density of matter available for neutrino interaction.

\begin{figure}[h]
\begin{center}
\mbox{\epsfig{file=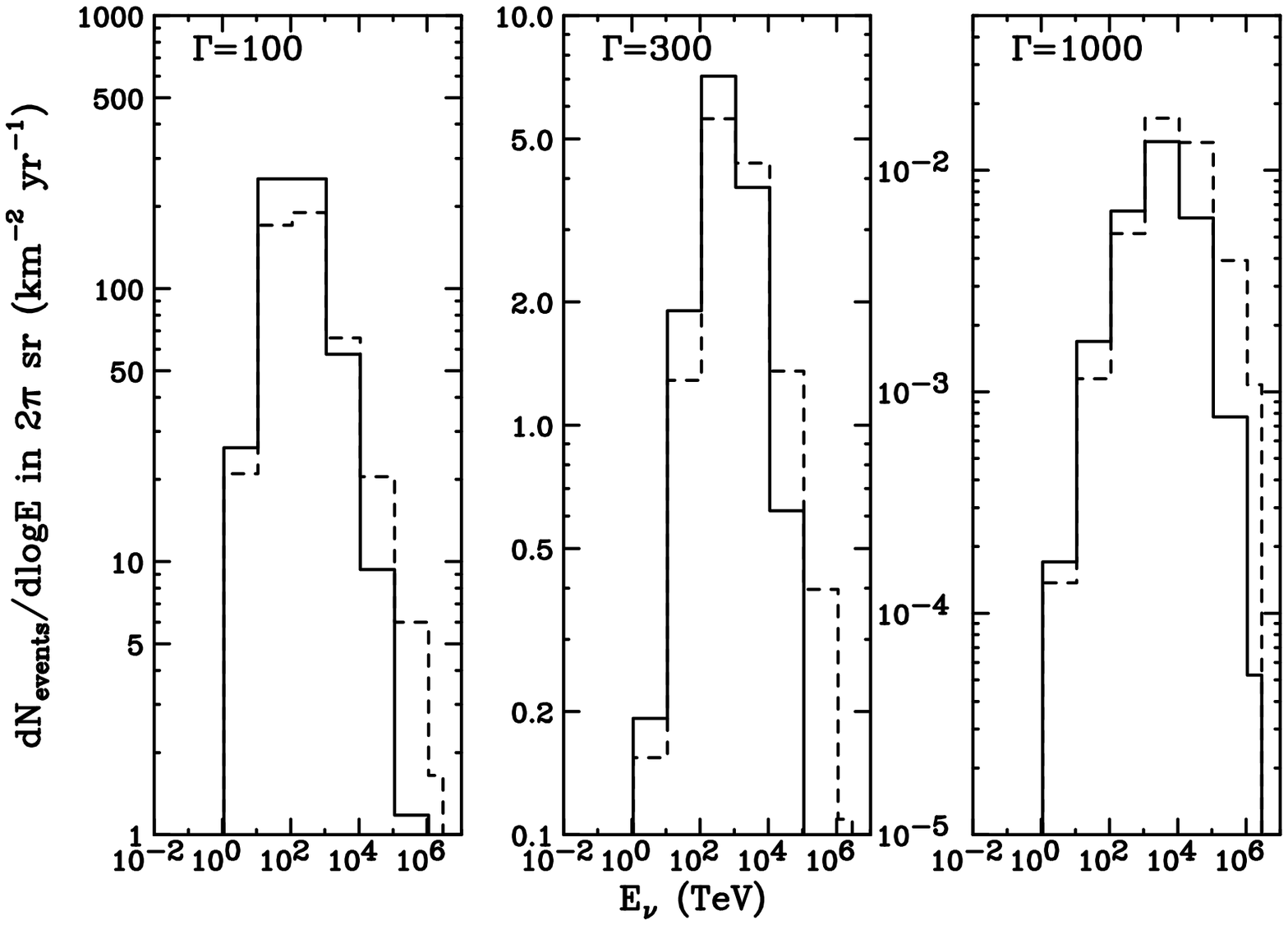,height=10cm}}
\end{center}
\caption{Energy distribution of $\nu_\mu+\bar\nu_\mu$ events
for $\Gamma=$100, 300 and 1000. Solid (dashed) lines correspond to
upgoing (downgoing) neutrinos. Only fluctuations in distance and
energy are accounted for. Absorption in the Earth and the limited target above 
the detector are taken into account.}
\label{energydist}
\end{figure}

\begin{figure}[h]
\begin{center}
\mbox{\epsfig{file=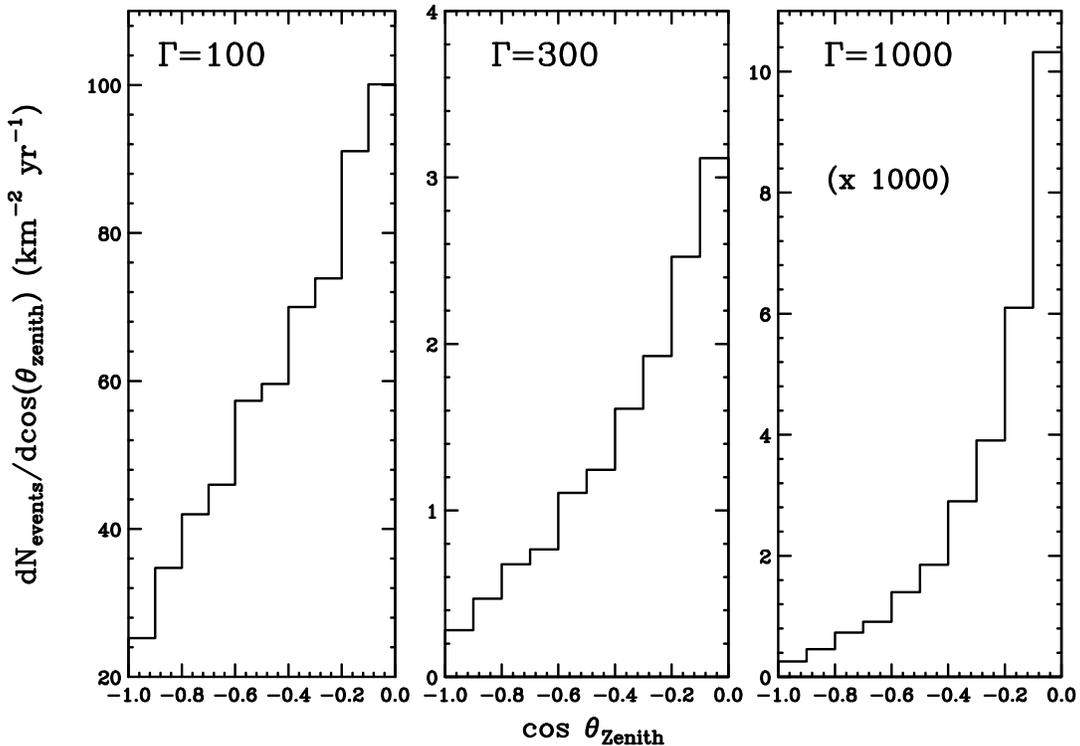,height=10cm}}
\end{center}
\caption{$\cos(\theta_{\rm zenith})$ distribution of $\nu_\mu+\bar\nu_\mu$
events for $\Gamma=$100, 300 and 1000. $\cos(\theta_{\rm zenith})=-1$
corresponds to upgoing neutrinos and $\cos(\theta_{\rm zenith})=0$
to horizontal. Notice that for $\Gamma=1000$ we have multiplied the
event rate by a factor 1000.}
\label{zenithdist}
\end{figure}

In Figure \ref{energydist}, we show the energy dependence of upgoing and
downgoing $\nu_\mu+\bar\nu_\mu$ events for three representative
values of $\Gamma$.
The zenith angle distribution of upgoing neutrinos is shown in Figure
\ref{zenithdist}, i.e. $-1<\cos(\theta_{\rm zenith})<0$. As
$\Gamma$ increases, the higher energy neutrinos are
attenuated by the Earth. This explains why, as $\Gamma$ increases, the
distributions increasingly resemble an exponential attenuation function.

\subsection{$\nu_\tau+\bar\nu_\tau$ events}

$\nu_\tau$ production is expected to be
very small in GRBs and in general in any astrophysical environment
where $\nu$'s are produced
in $p -p$ or $p - \gamma$ collisions. Several calculations suggest a ratio
\cite{athar}:
\begin{equation}
F_{{\nu_\tau}/{\nu_\mu}}=
{\Phi(\nu_\tau+\bar\nu_\tau)\over \Phi(\nu_\mu+\bar\nu_\mu)}\sim 10^{-5}~,
\end{equation}

\begin{figure}[h]
\begin{center}
\mbox{\epsfig{file=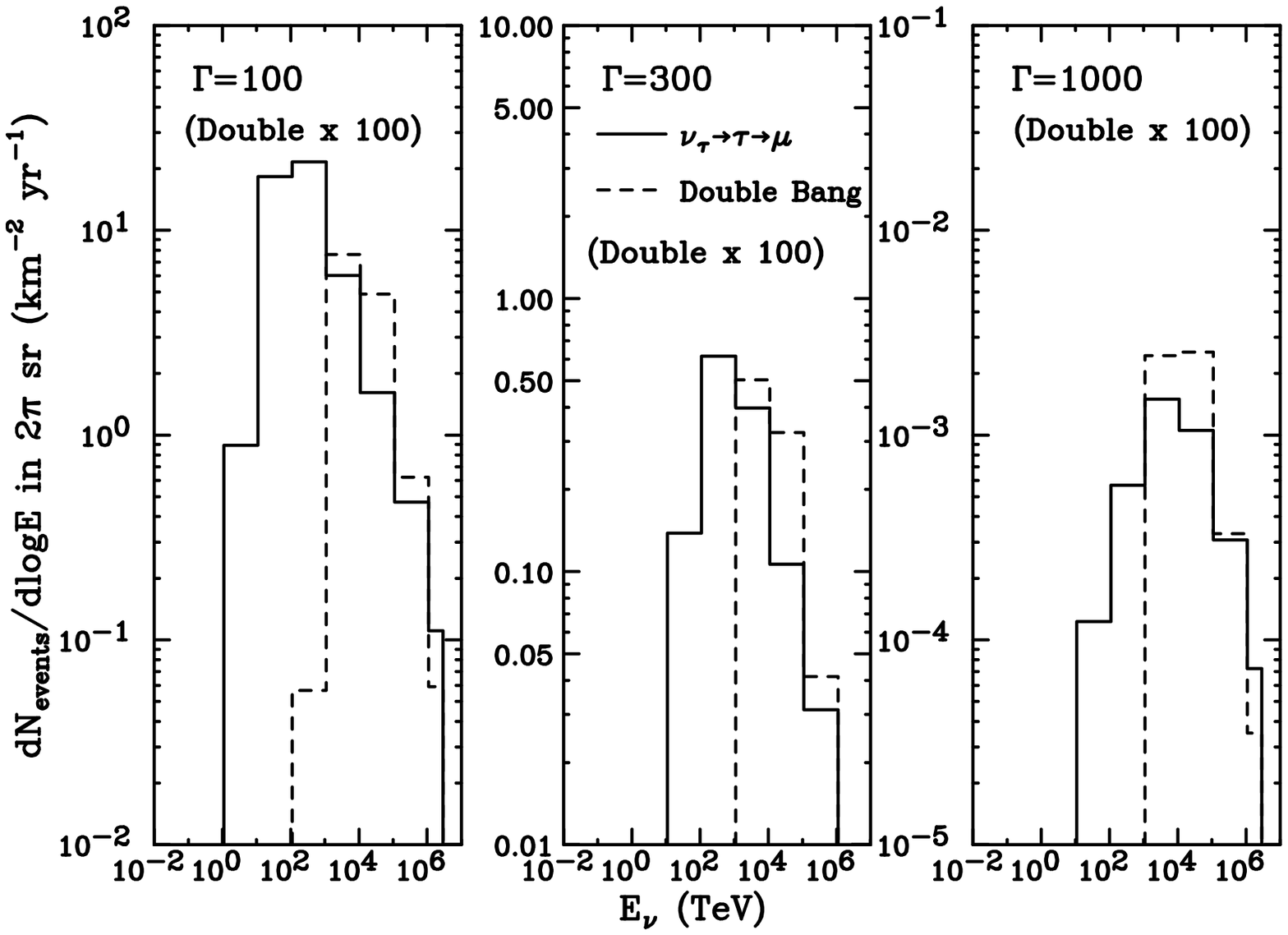,height=10cm}}
\end{center}
\caption{Energy distribution of upgoing
$\nu_\tau+\bar\nu_\tau$ double bang events (dashed line)
and events in which the $\tau$ decays to $\mu$ (solid line)
for $\Gamma=$100, 300 and 1000.
The double bang event rate has been multiplied
by a factor 100 in all the plots. Absorption in the Earth and the
limited target above the detector are taken into account.}
\label{energydisttau}
\end{figure}

Oscillation scenarios in which $\nu_\mu$'s convert into
$\nu_\tau$'s can however provide abundant sources of $\nu_\tau$'s. Assuming
a typical value $F_{{\nu_\tau}/{\nu_\mu}}=0.5$ suggested by
SuperKamiokande measurements \cite{superk},
we obtain the double bang \cite{learned}
\footnote{Events in which two separated showers can be identified,
one initiated by the struck nucleon and the other by the decay of
the $\tau$ produced in the $\nu_\tau$ charged current interaction.}
event rates shown in Table \ref{nutauevents}.
The probability of detecting a $\nu_\tau$ induced
double bang event in IceCube is typically two orders of
magnitude smaller than the probability of observing a $\nu_\mu$ at
energies around 10 PeV \cite{athar,halzentau}.
This explains the small values of the event rates in
Table \ref{nutauevents}. In the event rate calculation we have accounted for
the energy loss of the $\nu_\tau$'s when they propagate along
the Earth's interior. This produces a pileup of events around
100 TeV, as pointed out in \cite{saltzberg}, reducing the number
of upgoing double bang events with respect to the downgoing ones.
This is due to the probability of $\nu_\tau$-induced
double bang detection which is
limited to a broad peak between $\sim$ 1 PeV and $\sim$ 100 PeV
(outside of which it is
negligible). This also explains the fact that the
energy distribution of the event rate peaks in the vicinity of 10 PeV
(see Figure \ref{energydisttau}).

We also calculated the $\nu_\tau+\bar\nu_\tau$ events that would be detected
by the appearance of a $\tau$ which decays to $\mu$ just below the
detector. The event rates are also shown in Table \ref{nutauevents}.
In this case the energy distribution of the events peaks around the
energy at which the events pileup (100 TeV);
see Figure \ref{energydisttau}. The probability of
detecting the $\mu$ is $\sim~17~\%$ of the probability of detecting
it in a $\nu_\mu$ interaction due to the branching ratio of the
$\tau$ to $\mu$ decay channel. This accounts for
a factor $\sim$ 6 difference
between $\nu_\tau\rightarrow\tau\rightarrow\mu$ in Table \ref{nutauevents} and 
the $\nu_\mu+\bar\nu_\mu$ rate in Table \ref{numuevents}.

\begin{table}
\begin{center}
\begin{tabular}{|c|c|c||c|c|} \hline
 & \multicolumn{2}{c||}{~Double Bang~} &  
\multicolumn{2}{c|}{~$\nu_\tau\rightarrow \tau\rightarrow \mu$~} \\ \cline{2-5}  

~{\raisebox{2.5ex}[0pt]{~Events/(${\rm km^2}$~yr) in $2\pi$  
sr}}~~&~Downgoing~~&~Upgoing~~&~Downgoing~~&~Upgoing~~\\ \hline\hline
$\Gamma=100$~~&~0.54~&~0.13~&~38~&~49~\\ \hline
$\Gamma=300$~~&~$3.6\times 10^{-2}$~&~$8.8\times 10^{-3}$~&~1.2~&~1.3~\\ \hline
$\Gamma=1000$~~&~$2.9\times 10^{-4}$~&~$6.6\times 10^{-5}$~&~$3.6\times  
10^{-3}$~&~$5.5\times 10^{-3}$~\\ \hline
\end{tabular}
\end{center}
\caption[] {$\nu_\tau+\bar\nu_\tau$ double bang events and
events in which the $\tau$ decays to $\mu$ $(\rm
km^{-2}~yr^{-1})$. Only fluctuations in distance and energy are considered.
Absorption in the Earth and the limited column of matter above the detector
are taken into account.
\label{nutauevents}}
\end{table}

\begin{figure}[h]
\begin{center}
\mbox{\epsfig{file=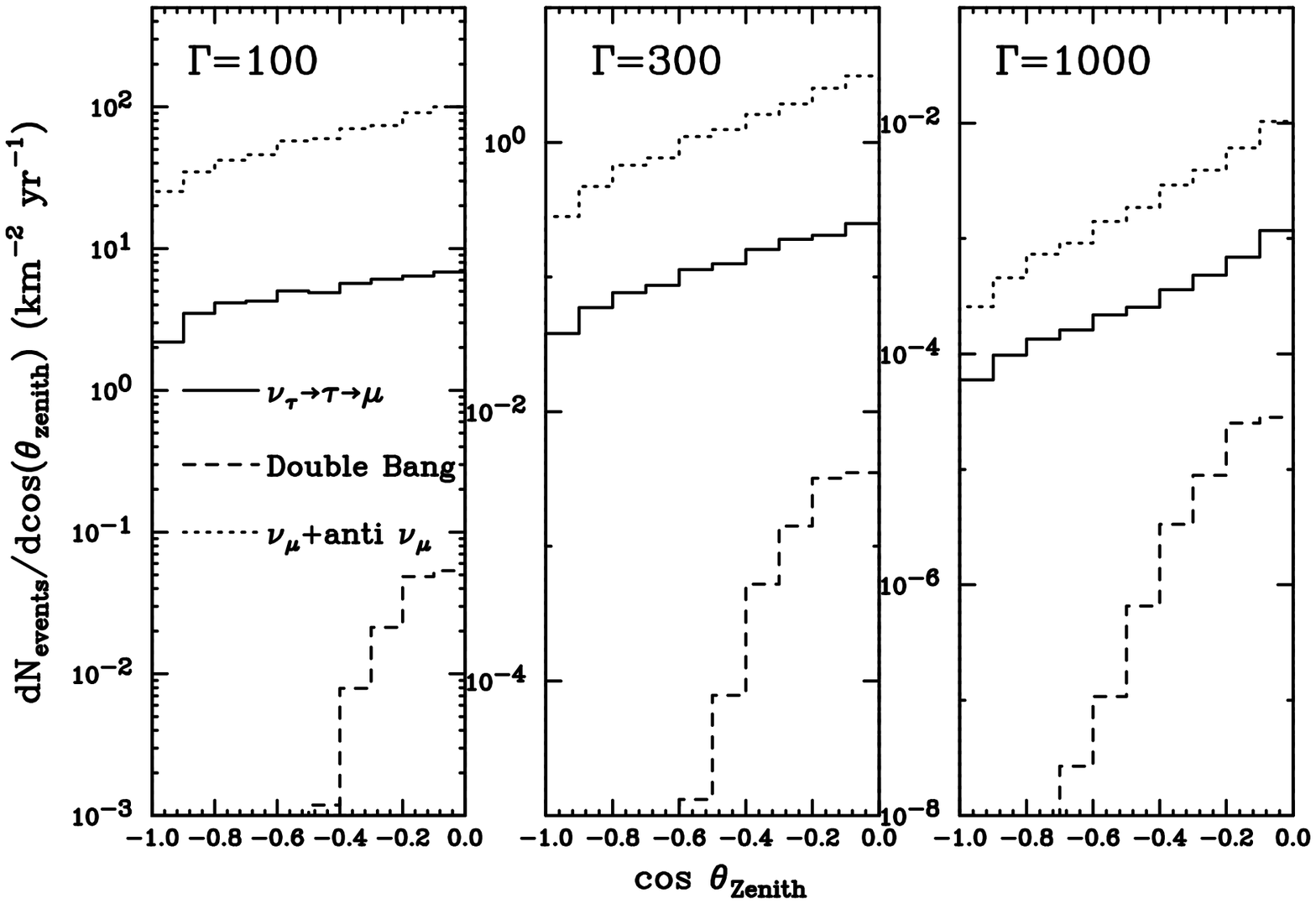,height=10cm}}
\end{center}
\caption{$\cos(\theta_{\rm zenith})$ distribution of $\nu_\tau+\bar\nu_\tau$
double bang events (dashed line)
as well as events in which the $\tau$ decays to $\mu$ (solid line),
for $\Gamma=$100, 300 and 1000. $\cos(\theta_{\rm zenith})=-1$
corresponds to upgoing neutrinos and $\cos(\theta_{\rm zenith})=0$
to horizontal. Also shown for comparison
are the $\nu_\mu+\bar\nu_\mu$ event rates (dotted line).}
\label{zenithdisttau}
\end{figure}

The expected event rates generated by both mechanisms
(double bang and $\tau\rightarrow\mu$ decay)
have very different and characteristic zenith angle distributions.
These are shown in Figure \ref{zenithdisttau}.
For large column depths
inside the Earth (i.e. $\cos(\theta_{\rm zenith})\sim -1$), the
events pileup around 100 TeV and the small probability of
detecting a double bang event at this energy reduces the number of
double bangs with respect to events in which a $\mu$ is detected.
When $\cos(\theta_{\rm zenith})\sim 0$, because the amount of matter
neutrinos have to cross is very small, there is no pileup of events
and the angular distributions have roughly the same shape.
Figure \ref{zenithdisttau} also shows the zenith angle distribution of
events produced by $\nu_\mu+\bar\nu_\mu$ for comparison, making it clear
that despite of the flatter distribution of the $\mu$'s from
$\nu_\tau+\bar\nu_\tau$ they are still outnumbered by $\mu$'s from
$\nu_\mu$ interactions.

\subsection{Fluctuations in $\Gamma$}

\begin{figure}[h]
\begin{center}
\mbox{\epsfig{file=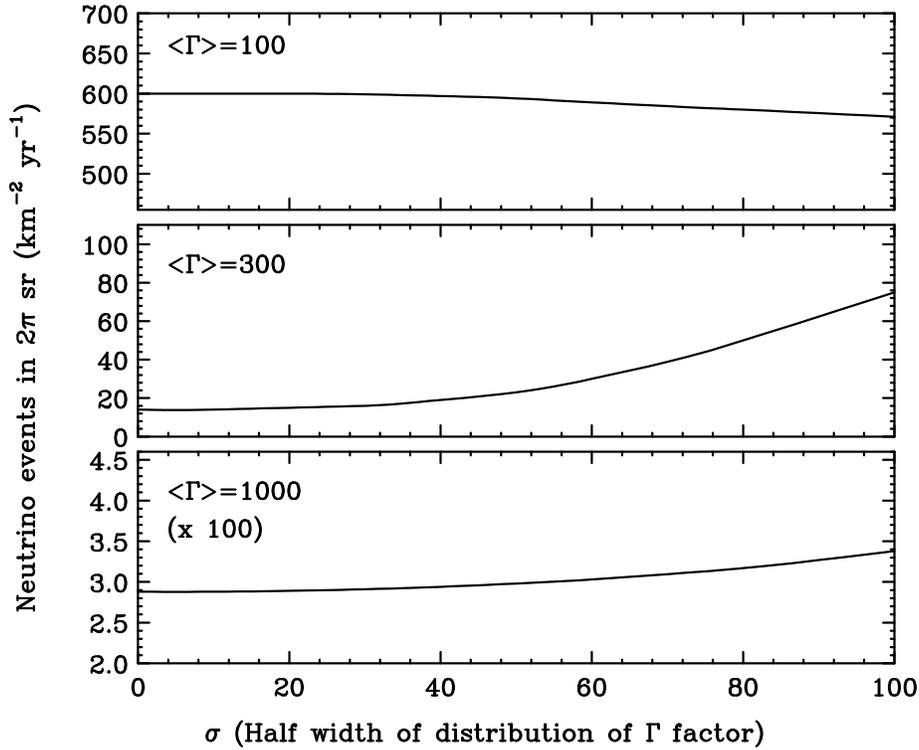,height=10cm}}
\end{center}
\caption{Upgoing $\nu_\mu+\bar\nu_\mu$ event rate as a function
of $\sigma$, the half width of the boost distribution around
$\left<\Gamma\right>$, for $\left<\Gamma\right>$=100, 300 and 1000. The rate
for $\left<\Gamma\right>=1000$ has been scaled up by a factor 100.
Absorption in the Earth is taken into account.}
\label{ratesigma}
\end{figure}

Finally, we discuss burst to burst fluctuations in the boost factor
$\Gamma$. So far we, quite unrealistically, assumed that $\Gamma$ is
single valued. We have, at present, no information on the distribution of
$\Gamma$ factors. It has been shown \cite{hooper} that the neutrino rates
can be significantly enhanced by fluctuations around the average value. It
may be, of course, that the fluctuations in energy and distance
which we took into
account following the experimental evidence, already reflect some or all
of the fluctuations in the boost factor. Following reference \cite{hooper}
we will illustrate the effect by assuming Gaussian distributions with
half width $\sigma$. The results are shown in Table \ref{numugammaave} for
different values of $\left<\Gamma\right>$. It is interesting to note that for
$\left<\Gamma\right>=100$ the event rates are almost independent of the value
of $\sigma$. For $\Gamma$ around 100 the
event rate is weakly dependent on $\Gamma$ due to the
saturation of the amount of energy that goes into neutrino production
(see Figures \ref{grbfluxes} and \ref{grbgamma}) and hence fluctuations in the 
value of the boost factor do not affect the event rate. This is not
the case when $\left<\Gamma\right>=300$ or 1000 since the event rate behaves  
roughly as
$\Gamma^{-4}$ in that $\Gamma$ range. Figure \ref{ratesigma} shows
more clearly the dependence of the event rate on $\sigma$ for
different values of $\left<\Gamma\right>$ illustrating these points.
It can be easily shown that when $\sigma \ll  \left<\Gamma\right>$
the following relation holds:
\begin{equation}
{{\rm Rate}(\left<\Gamma\right>\pm\sigma)\over {\rm Rate}(\left<\Gamma\right>)}
\propto {\sigma\over\left<\Gamma\right>}~,
\end{equation}
explaining why for $\left<\Gamma\right>=1000$ the variation of the event rate
with $\sigma$ is not as strong as when
$\left<\Gamma\right>=300$, even though the rate scales with $\Gamma^{-4}$ in
both cases. The same comments apply to double bang events produced
by $\nu_\tau$'s and events in which the produced $\tau$ decays to
$\mu$ (both are shown in Table \ref{nutaugammaave}).

\begin{table}
\begin{center}
\begin{tabular}{|c|c|c|c|} \hline
\multicolumn{2}{|c|}{} & \multicolumn{2}{c|}{~Absorption~} \\ \cline{3-4}
\multicolumn{2}{|c|}{\raisebox{2.5ex}[0pt]{~Events/(${\rm km^2}$~yr) in  
$2\pi$~sr}}&~Downgoing~~&~Upgoing~~\\ \hline\hline
&~$\sigma=0$~&~476~&~600~\\ \cline{2-4}
&~$\sigma=30$~&~476~&~600~\\ \cline{2-4}
$\left<\Gamma\right>=100$~~&~$\sigma=50$~&~472~&~594~\\ \cline{2-4}
&~$\sigma=75$~&~464~&~582~\\ \cline{2-4}
&~$\sigma=100$~&~454~&~571~\\ \hline\hline
&~$\sigma=0$~&~13~&~14~\\ \cline{2-4}
&~$\sigma=30$~~&~15~&~16~\\ \cline{2-4}
$\left<\Gamma\right>=300$~~&~$\sigma=50$~~&~22~&~23~\\ \cline{2-4}
&~$\sigma=75$~~&~39~&~44~\\ \cline{2-4}
&~$\sigma=100$~~&~63~&~75~\\ \hline\hline
&~$\sigma=0$~&~$4.2\times 10^{-2}$~&~$2.8\times 10^{-2}$~\\ \cline{2-4}
&~$\sigma=30$~~&~$4.2\times 10^{-2}$~&~$2.9\times 10^{-2}$~\\ \cline{2-4}
$\left<\Gamma\right>=1000$~~&~$\sigma=50$~~&~$4.3\times 10^{-2}$~&~$3\times  
10^{-2}$~\\ \cline{2-4}
&~$\sigma=75$~~&~$4.5\times 10^{-2}$~&~$3.1\times 10^{-2}$~\\ \cline{2-4}
&~$\sigma=100$~~&~$4.8\times 10^{-2}$~&~$3.4\times 10^{-2}$~\\ \hline\hline
\end{tabular}
\end{center}
\caption[] {$\nu_\mu+\bar\nu_\mu$
events $(\rm km^{-2}~yr^{-1})$, taking into account
fluctuations in boost factor $\Gamma$, distance and energy.
The distribution of boost factors is assumed to be
a Gaussian of half width $\sigma$ centered in $\left<\Gamma\right>$.
Absorption in the Earth and the limited column of matter available
for neutrino interaction above the detector are taken into account.
\label{numugammaave}}
\end{table}

\begin{table}
\begin{center}
\begin{tabular}{|c|c|c|c||c|c|} \hline
\multicolumn{2}{|c|}{} & \multicolumn{2}{c||}{~Double Bang~} &  
\multicolumn{2}{c|}{~$\nu_\tau\rightarrow \tau\rightarrow \mu$~} \\ \cline{3-6}
\multicolumn{2}{|c|}{\raisebox{2.5ex}[0pt]{~Events/(${\rm km^2}$~yr) in $2\pi$  
sr~}}&~Downgoing~~&~Upgoing~~&~Downgoing~~&~Upgoing~~\\ \hline\hline
&~$\sigma=0$~&~0.54~&~0.13~&~38~&~49~\\ \cline{2-6}
&~$\sigma=30$~&~0.52~&~0.12~&~37~&~48~\\ \cline{2-6}
$\left<\Gamma\right>=100$~~&~$\sigma=50$~&~0.48~&~0.11~&~36~&~46~\\ \cline{2-6}
&~$\sigma=70$~&~0.43~&~0.1~&~33~&~43~\\ \cline{2-6}
&~$\sigma=100$~&~0.39~&~0.09~&~32~&~40~\\ \hline\hline
&~$\sigma=0$~&~$3.6\times 10^{-2}$~&~$8.8\times 10^{-3}$~&~1.2~&~1.3~\\ \cline{2-6}
&~$\sigma=30$~~&~$4.0\times 10^{-2}$~&~$1.0\times 10^{-2}$~&~1.4~&~1.5~\\ \cline{2-6}
$\left<\Gamma\right>=300$~~&~$\sigma=50$~~&~$5.0\times 10^{-2}$~&~$1.2\times  
10^{-2}$~&~1.9~&~2.1~\\ \cline{2-6}
&~$\sigma=70$~~&~$6.9\times 10^{-2}$~&~$1.6\times 10^{-2}$~&~2.9~&~3.3~\\ \cline{2-6}
&~$\sigma=100$~~&~$9.9\times 10^{-2}$~&~$2.4\times 10^{-2}$~&~5.0~&~6.2~\\  
\hline\hline
\end{tabular}
\end{center}
\caption[] { 
$\nu_\tau+\bar\nu_\tau$ double bang events and events in which the $\tau$
decays producing a $\mu$
$(\rm km^{-2}~yr^{-1})$. We take into account
fluctuations in boost factor $\Gamma$, distance and energy.
The distribution of boost factors is assumed to be
a Gaussian of half width $\sigma$ centered in $\left<\Gamma\right>$.
Absorption in the Earth and the limited column of matter available
for neutrino interaction above the detector are taken into account.
\label{nutaugammaave}}
\end{table}
%
%

\section{Summary and conclusions}

We have investigated potential signatures in ${\rm km^3}$ telescopes
of high energy neutrino fluxes produced in $p-\gamma$ interactions in
GRB environments. We stress the fact that the rate is dominated by
fluctuations in distance, GRB energy and in
the bulk Lorentz factor $\Gamma$ of the
expanding GRB fireball. We have used an euclidean distribution
to account for fluctuations in distance.
Using a distribution in which GRB's follow the star formation rate,
as suggested in \cite{mao}, the event rates are reduced only by $\sim 20\%$.
On the other hand, a cosmological distribution following galaxies
\cite{peebles,mannheim}
allows for more close GRB's so that the event rate increases
by a factor of 20 (see Table \ref{grbdist}). Neutrino telescopes
may help to constrain the distribution in distance of GRBs.
The most relevant parameter is $\Gamma$, which
determines the rate of $p -\gamma$ interactions and hence the amount
of energy that goes into neutrino production.
For small values of $\Gamma$ ($\Gamma\sim 50$ or less) the expansion of
the GRB fireball is not
sufficiently fast and the large photon density makes it opaque to
$p -\gamma$, efficiently producing pions --- the parents of the neutrinos ---
and saturating the amount of energy available for neutrino production.
Due to the scaling of the energy break in the spectrum with $\Gamma^2$,
mostly low energy neutrinos are produced
whose detection efficiency is smaller
due to the small range of the muon at low energies
and the relatively high energy threshold of the neutrino
telescopes ($\sim 100$ GeV). However the large neutrino flux compensates
for the small detection efficiency.
For large values of $\Gamma$ ($\Gamma>300$), an increasingly
larger fraction of the neutrino energy goes into the high
energy part of the spectrum, however the overall amount of energy
is very small, producing small event rates.
Values of $\Gamma<100$ give rise to
large number of events that would even be observable in existing
smaller neutrino telescopes such as AMANDA
\cite{AMANDA} (one has
to scale the results down by roughly two orders of magnitude
to account for the
smaller effective area of the detector).

We have shown that absorption of the upgoing $\nu_\mu$'s
inside the Earth, as well as the
limited column of matter available for
downgoing neutrino interactions, play a relevant
role, making upgoing and downgoing event rates roughly equal. The change
of the energy break of the spectrum with $\Gamma^2$ combined
with the absorption of the Earth is reflected in
the zenith angle distributions of the event rates
which may give some complementary information about $\Gamma$.

GRB neutrino detection with
$\rm km^3$ neutrino telescopes also has the potential to investigate
$\nu_\mu\rightarrow\nu_\tau$ oscillations over a baseline of 1000 Mpc.
Double bang $\nu_\tau$ induced events offer an unmistakeable
signature which allows downgoing $\nu_\tau$ detection.
A $\rm km^3$ telescope operating for 10 years may detect $\sim$ 10
downgoing double bang events if $\Gamma=100$ without any
potential background.
Upgoing double bang events are not going to be detected since the
$\nu_\tau$'s pileup around 100 TeV, where the probability of a double
bang detection is negligible. Muons produced in $\tau$ decays
are outnumbered by $\mu$'s from $\nu_\mu$ interactions at least for the
type of fluxes expected from GRBs \cite{halzentau}; besides,
their energy distribution does not show a clear
and characteristic signature so they are difficult to identify.

In summary, neutrino telescopes open up the possibility of determining the value
of $\Gamma$ and its fluctuations, as well as the possibility of identifying
$\nu_\tau$'s. They are powerful instruments to reveal important astrophysical
information about the most energetic objects ever observed in the universe
and about neutrino oscillation scenarios over cosmological baselines.

\begin{table}
\begin{center}
\begin{tabular}{|c|c|c|c|} \hline
{~Events/(${\rm km^2}$ yr) in $2\pi$ sr~}&~Euclidean~&~Cosmological~&~Star  
Formation Rate~~\\ \hline\hline
~$\Gamma=100$~&~1076~&~21,029~&~832~ \\ \hline
~$\Gamma=300$~&~27~&~424~&~20~ \\ \hline
~$\Gamma=1000$~&~0.3~&~1.8~&~$6 \times 10^{-2}$~ \\ \hline
\end{tabular}
\end{center}
\caption[] {$\nu_\mu+\bar\nu_\mu$
events $(\rm km^{-2}~yr^{-1})$. Only fluctuations in
distance and energy are taken into account. Different
distributions in distance are used: the first column corresponds
to an euclidean distribution, the second to a cosmological distribution
following galaxies \cite{peebles}
and the third one assumes that GRB's follow the
distribution of star formation regions \cite{mao}.
Absorption in the Earth as well as the limited column of matter above
the detector are taken into account. 
\label{grbdist}}
\end{table}
%
%

\section*{Acknowledgments}
This research was supported in part by the US Department of Energy under
grant DE-FG02-95ER40896 and in part by the University of Wisconsin
Research Committee with funds granted by the Wisconsin Alumni Research
Foundation.
J.A. thanks the Department of Physics, University of Wisconsin, Madison and the
Fundaci\'on Caixa Galicia for financial support.

\end{document}